\documentclass[12pt]{article}
\usepackage{graphicx}
\usepackage{bm}

\newcommand{\beq}{\begin{equation}}
\newcommand{\eeq}{\end{equation}}

\begin{document}

\title{QBC3: An Unconditionally Secure Quantum Bit Commitment Protocol}

\author{Horace P.~Yuen\\Department of Electrical Engineering and Computer Science, \\
Department of Physics and
Astronomy, \\ Northwestern University, Evanston, IL, 60208, \\
Email: yuen@eecs.northwestern.edu} \maketitle

\begin{abstract} This article describes a quantum bit commitment protocol, QBC3, based on entanglement
destruction via forced measurements and proves its unconditional
security. Some comments on the current status of the field are
also made.
\end{abstract}

\section{Introduction}
It is nearly universally accepted that unconditionally secure
quantum bit commitment (QBC) is impossible. For a summary of the
problem, its brief history and further references, see
\cite{yuen03,dariano06}. In the following, my current view of the
field is summarized and a new secure protocol, QBC3, is presented
together with a sketch of the security proof.

\section{The Impossibility Proof}
The general claim of impossibility was made in ref.
\cite{mayers97,lochau} and criticized in some detail in
\cite{yuen03} and \cite{yuen02}. In my current understanding, there
are two major gaps in the arguments supporting such claims. The
first is the assumption that all use of (classical) randomness in
the protocol can be purified via quantum entanglement that results
in an openly known pure state for the parties A and B. The second is
that any quantum measurement specified in the middle of protocol
execution can be postponed to the end of commitment just before
opening. These two points are related, and both connected to the
implicit assumption that entanglement can be formed and maintained
to just before opening.

The first assumption is clearly false by noting that entanglement
has to end at some point, at which the entanglement basis and
probabilities can still be determined by random numbers. Such
numbers may be physically generated classically or quantum
mechanically, say by throwing a die or heterodyning the vacuum.
They are needed in many classical and all quantum cryptographic
protocols. There is no need for the basis and probabilities to be
known to the other party if his/her security is not affected -- we
have codified this as the Secrecy Principle \cite{yuen02,yuen03}.
In particular, this does not violate the Kerckhoff's Principle,
which states that the adversary is presumed to know the structure
of the protocol in cryptography. On the other hand, the
"community" somehow thinks there is a violation according to ref.
\cite{dariano06}.

Similar to the first, the second assumption is true in various
types of protocols but not all. The crucial point is whether the
relevant entanglement that is responsible for successful cheating
by either A or B can be created or can remain effective after
measurement in the middle of a protocol.

I have insisted, since the beginning in 2000, that a priori there
can be no general impossibility proof without mathematically
representing a common characteristic of all QBC protocols, which has
never been given. As Ozawa put it, nobody calls the Church-Turing
thesis the Church-Turing Conjecture or Theorem, because there is no
mathematical definition of a mechanical procedure. It is difficult
to understand on what basis one can claim to have a completely
general impossibility theorem, as allegedly maintained by the vast
majority of the `community' at present \cite{dariano06}. Apparently,
much remains to be written to explain just this point.

\section{Exploiting Gaps in the Impossibility Proof}
There have been many attempts by different authors to produce
protocols that lie outside the impossibility proof formulation.
Even when successful, a secure protocol has not been created
except my QBC1 \cite{yuen03}. My own attempts are summarized in
Appendix A of \cite{yuen03} and will be updated in the future. One
of the reasons that error is easily committed in the QBC problem
is conceptual -- how one should represent certain given physical
action mathematically. Indeed, I believe this act/math
correspondence is a fruitful area for future exploration that
would not only lead to a more systematic understanding of all
possible protocol formulations and attacks in QBC, but in the
foundation of quantum physics generally.

More general proofs of impossibility relaxing the first assumption
have been given by Ozawa \cite{ozawa} and Cheung \cite{cheung}
independently, and also by D`Ariano etc \cite{dariano06} in an
algebraic formulation that has not been translated into the
standard form. While it is claimed in \cite{dariano06} that their
formulation covers all my previous protocols, it is not true for
QBC1.

\section{Protocol QBC3}
My current QBC3 is similar to a previous QBC3 I had that is not
secure, its security now derives from an added feature from B's
checking before opening which forces A to make a measurement that
destroys his cheating entanglement. Consider the following
protocol: B sends A a sequence of $n$ qubits, each randomly in one
of the four BB84 states $|j_{l}\rangle$, $j_{e} \in \{1,2,3,4\}$,
named by its position in the sequence. A randomly picks one,
modulates it by  $U_{0}=R(\pi/16)$ or $U_{1}=R(-\pi/16)$, rotation
by $\pm\pi/16$ on the great circle containing $\{|j\rangle\}$
depending on $b \in \{0,1\}$, and sends it back to B as
commitment. A opens by sending back the rest and revealing
everything. This is a typical preliminary protocol in that it is
at best $\epsilon$-concealing with A's optimum cheating
probability not correspondingly close to 1, so that in a sequence
of such action a single bit can be committed that is both
$\epsilon$-concealing and $\epsilon$-binding. For definitions and
terminology, see \cite{yuen03}.

This protocol was first given as an anonymous state protocol in
which B entangles her randomness with arbitrary probability and
basis, but it was found, as asserted above, that A's cheating
transformation is actually independent of such information.  There
are many ways for A to entangle and it is claimed in
\cite{dariano06} that the one with the minimum ancilla state-space
dimension is as good as any \cite{dariano06}. Let $|l\rangle \in
H_{A}$ be the entanglement ancilla states, $P$ the cyclic shift
unitary operator on $n$ qubits, $P^n=I$. Let us first assume A
entangles in this minimal way,
\begin{equation}\label{1}
    |\Psi_{b}\rangle=U_{b}\frac{1}{\sqrt{n}}\sum_{\ell=1}^n|\ell\rangle\otimes
    P^{\ell}|j_{1}\rangle ...|j_{n}\rangle
\end{equation}
where $|j_{1}\rangle$ is acted on by $U_{b}$.  The protocol can be
shown to be $\epsilon$ - concealing similar to the proof given in
ref \cite{yuen01}, and A can locally turn $|\Psi_{0}\rangle$ to
$|\Psi_{1}\rangle$ near perfectly.

Consider the following addition to the protocol.  Before opening,
B asks A to send back a fraction $\lambda$, say
$\lambda=\frac{1}{2}$, of the $n$ qubits chosen randomly by B for
checking. One can choose $n$ large enough so that $\frac{n}{2}$ is
long enough for the $\epsilon$ concealing level. If A claims that
fraction contains the committed one, B would ask to check the
remaining $1-\lambda$ fraction instead. Assuming, as usual, that A
must open one bit value perfectly, A would have to answer B's
checks perfectly. A continuity argument would take care of the
general case. The best he can do is to perform a Luders
measurement with controlled swaps on his ancilla $H_{A}$,
projecting into the qubits B picked. If the checking qubits do not
contain the committed qubit from the measurement result, which is
then still entangled with the rest, A succeeds in cheating.  If
they do, A would have disentangled the committed qubit from the
rest, and entanglement cheating for A is no longer possible.
Indeed, his cheating success probability is not small, given by
$1/2$ for $\lambda = 1/2$, while the protocol remains
$\epsilon-$concealing. This already contradicts the quantitative
conclusion of the Impossibility Proof, for which A's cheating
probability should be close to 1. The protocol is extended to be a
near-perfect concealing and binding one, QBC3, as described
generally in \cite{yuen03} and \cite{yuen00}, by adjusting the
following $m \rightarrow \infty$ with $n/m \rightarrow \infty$.
\begin{center} \vskip 0.1in \framebox {
\begin{minipage}{0.9\columnwidth}
\vskip 0.1in \underline{PROTOCOL {\bf QBC3}}

{\small \begin{enumerate}
\item B sends A $m$ segments of $n$-qubits, each randomly in one of four BBS4 states.
\item A forms (1) from each segment, modulates the first qubit by the same
$U_{b}=R(\pm\pi/16)$ and sends back to B these first qubits.
\item B randomly chooses half of the qubits in each segment and asks A to send them back for checking.
If for any segment A claims it contains the committed one, B asks
to check the other half instead.
\item A opens by sending back all qubits and revealing everything; B
verifies.
\end{enumerate}
\vskip 0.1in }
\end{minipage}}
\end{center}
\vskip 0.11in

In the usual Impossibility Proof formulation, an honesty
assumption is made for both parties on their entanglement/state
formation in multi-pass protocols. Thus, the above protocol
\emph{already} shows such proofs are \emph{not} correct, even
though A can actually cheat with other entanglements than (1).

For checking before commitment, two ways were proposed before to
deal with it quantitatively: by ensemble checking or by a
classical game-theoretic formulation -- see the Appendix of ref
\cite{yuen03b}.  It is clear that some decision has to made on
what to do if A or B is found cheating, they cannot be allowed to
go free or they could just keep cheating until they succeed.  In
the ensemble checking formulation, the first time a party is found
cheating is taken as failure to cheat in the evaluation of their
cheating success probability. For the game theoretic formulation,
a penalty is assigned to each cheat detection and the total
penalty is bounded. Even a single random qubit can be checked by
requiring one party to entangle all the possibilities with the
other party checking the total state. Another approach is to add a
``holding phase'' \cite{dariano06} between commitment and opening,
during which cheating detection counts as failure to cheat. Such
an approach was used in several of my early checking protocols
\cite{yuen03,yuen00}.   Note that the formulation in ref
\cite{dariano06} does not include a checking action in the holding
phase, and it does not include any penalty on cheat detection.
Thus, it does not cover the present 2-pass protocol QBC3 or my old
3-pass QBC1. In particular, with (1) being enforced by A on the
qubits sent back by B, the security proof of QBC1 as outlined in
\cite{yuen03} is completed.

For QBC3, it is more complicated to check (1) since A already
committed, but it can be done as follows. A would first send his
ancilla to B with a random entanglement basis, then B sends back
A's committed qubit to him who would turn it back to neutral. That
B sends back the correct qubit can be checked by A via asking B to
send in all the states in her possession for A to verify. Then A
sends back all qubits to B who can now check (1). This sequence
can be carried out in either the ensemble or game-theoretic
formulation. In this way, a relatively simple but complete proof
has been given.

\section{Conclusion}

A protocol QBC3 with checking has been given that clearly lies
outside the formulation of all the impossibility proofs. Apart
from spelling out the details, all the essential elements for a
complete security proof for QBC3 have been described.

\section{Acknowledgement}
I would like to thank G. M. D'Ariano, R. Nair, M. Ozawa, M.
Raginsky, and especially C.Y. Cheung, for useful discussions.

\end{document}